## High-Energy Atmospheric Radiation: From Thunderstorm Ground Enhancements to Terrestrial Gamma-Ray Flashes

**A. Chilingarian**

**Yerevan Physics Institute, Alikhanyan Brothers 2, Yerevan, Armenia, AM0036**

**Abstract**

It is widely acknowledged that the atmosphere protects the Earth from harmful radiation; however, understanding its active role in particle acceleration is crucial for space weather and atmospheric physics research. This work presents a unified conceptual and observational framework, significantly enhancing our understanding by reinterpreting these radiation bursts as manifestations of the same runaway processes occurring at different atmospheric depths.
The Global Electric Circuit (GEC) manages enhanced particle fluxes from the atmosphere. Thunderstorms create strong atmospheric electric fields (AEFs) that extend over vast areas within and around storm systems. The energy contained in AEFs converts to radiation energy through various mechanisms, including relativistic runaway electron avalanches (RREAs, Gurevich et al., 1992), which appear as thunderstorm ground enhancements (TGEs, Chingarian et al., 2010, 2011) when detected on the Earth's surface and as gamma glows when observed above intense tropical thunderstorms. Exotic versions of these two primary radiation sources include brief gamma ray bursts recorded by orbiting gamma observatories at 500-700 km above the surface, namely terrestrial gamma ray flashes (TGDs, Fishman et al., 1994), and downward terrestrial gamma ray flashes (DTGFs, Ortberg et al., 2024), recorded in coincidence with lightning leaders. Despite their shared physical origin, these events are often treated separately due to differences in detection methods, duration, and altitude. This work presents a unified conceptual and observational framework that reinterprets these radiation bursts as manifestations of the same runaway processes occurring at distinct atmospheric depths. We discuss recent results from satellite (ASIM), aircraft (ALOFT), balloon (HELEN), and ground-based (SEVAN and KANAZAWA) experiments to demonstrate the advantages of the integrating approach. This study resolves key contradictions in the field, establishes new classification criteria based on physics rather than detector location, and reshapes our understanding of particle acceleration in thunderstorms.

**Key Points:**

- **Unified physical mechanism:** TGEs, gamma glows, TGFs, and DTGFs are all products of the RREA process occurring in different atmospheric regions.
- **Multi-platform data synthesis:** Includes results from ASIM (space), ALOFT (aircraft), HELEN (balloon), Kanazawa (DTGF), and SEVAN (ground).
- **Challenges to prevailing models:** Show that the lightning origin of seed electrons for TGFs and lightning leader-leader-based models of DTGF generation are inconsistent.

## 1. Introduction

Thunderstorms create strong electric fields that extend across large areas within and around the storm system. Charge separation in thunderclouds, driven by updrafts of warm air and interactions among hydrometeors, generates oppositely directed dipoles within the cloud.

C.T.R. Wilson, one of the first particle physicists and a pioneering researcher in atmospheric electricity, introduced a puzzling physical phenomenon known as “runaway” electrons at the beginning of the last century. This phenomenon involves the acceleration of electrons by strong electric fields in thunderclouds (see Chilingarian et al., 2025, and references therein). From 1945 to 1949, Joachim Kuettner conducted groundbreaking experiments at Zugspitze (Kuettner, 1950), where he discovered the tripole charge structure within thundercloud layers. According to the tripole model, the atmospheric electric field consists of upper and lower dipoles accelerating free electrons toward open space and the Earth's surface.

The upper dipole consists of the main negative and positive charge layers located in the middle and top of a thundercloud (see Fig.1). Electrons accelerated in this dipole create avalanches in the upper atmosphere, resulting in gamma-ray glows detected by airborne experiments flying above thunderstorms (Kelly et al., 2015; Ostgaard et al., 2019; Marisaldi et al., 2024; Helmerich et al., 2024). Occasionally, a few gamma rays reach orbiting gamma observatories 400-700 km away from the source, initiating microsecond bursts of particles known as terrestrial gamma flashes (TGFs, Fishman et al., 1994; Mailyan et al., 2015).

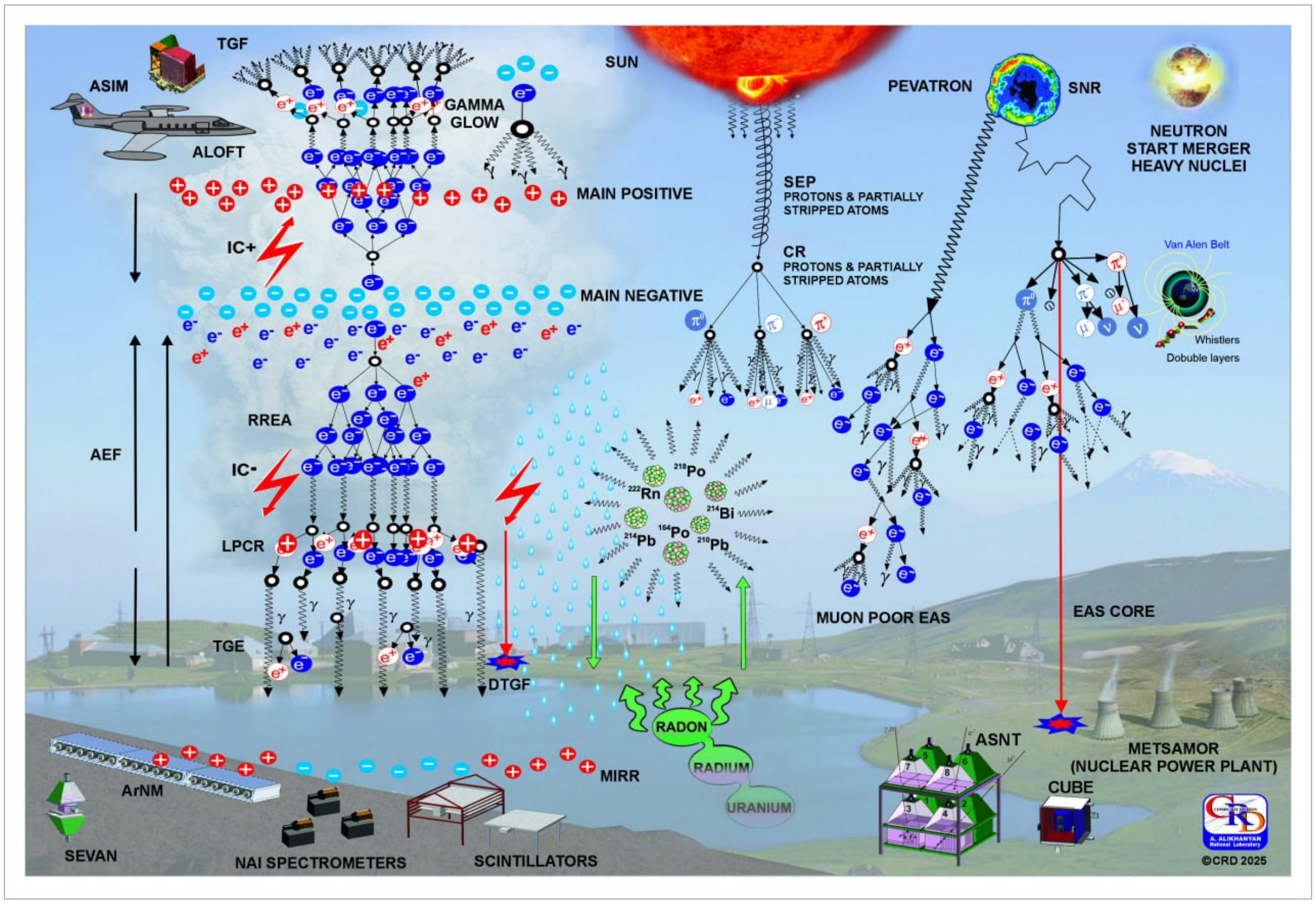

Figure 1. **The fluxes of secondary particles from space and atmospheric accelerators, as well as gamma radiation from $^{222}$Rn progeny. The cartoon also shows the charge structure of a thundercloud, the direction of electric fields, lightning flashes, and various measuring facilities on the surface and in space. Additionally, we can see the sources of primary cosmic rays. In the background is the Aragats cosmic ray station, located at 3200 meters and equipped with various particle detectors and spectrometers.**

The lower dipole comprises the same primary negative layer and an induced mirror charge on the Earth's surface. Occasionally, a third dipole forms between the main negative layer and a transient "pocket" of positively charged particles in the lower part of the cloud (LPCR), which is often associated with falling graupel (snow pellets coated with a layer of ice, Kuettner, 1950). Electrons accelerated by the two lower dipoles generate electron-gamma ray avalanches detected on the ground as thunderstorm ground enhancements (TGEs; Chilingarian et al., 2010, 2011). Due to the proximity of the electron accelerators in thunderclouds to the surface, TGEs contain millions of gamma rays, electrons, positrons, and neutrons. A fourth dipole between the LPCR and its mirror in the Earth accelerates positrons and positive muons while decelerating electrons and negative muons. Understanding these distinct charge structures and the hierarchy of particle acceleration mechanisms clarifies the high-energy phenomena observed in thunderstorms and their implications for atmospheric science.

The right side of Fig. 1 depicts the Extensive Air Showers (EASs, Auger et al., 1939) born in the high atmosphere by gamma rays and protons interacting with atmosphere atoms. High-energy particles are produced during violent explosions and mergers in the Galaxy and beyond. We show in the cartoon the supernova remnant and neutron star merger. However, there are also other sources of ultra-high energy particles; this field is actively investigated by new large experiments, reviewed in the next section. EASs, upon reaching the Earth's surface, can cover several square kilometers of area. The EAS cores contain the highest energy secondary particles, originating a very short, compact bursts shown by the red circle. Muon-poor events registered at altitudes above 4000 m indicate Pevatrons, the stellar sources enable protons to be accelerated up to $10^{15}$ eV. We also show the nuclear power plant, a possible source of radioactive contamination, and the Van Allen belt that can send MeV electrons toward the Earth's surface.

Particle detectors located beneath electron accelerators at high-altitude sites in Armenia, Eastern Europe, Germany, and other locations have recorded hundreds of TGEs, confirming extensive particle fluxes over thousands of square kilometers and durations ranging from seconds to tens of minutes (Chum et al., 2020; Chilingarian et al., 2024a). Data from Aragats have convincingly validated the RREA/TGE model through direct electron and gamma-ray spectrum measurements, which have also observed atmospheric neutrons and positrons (Chilingarian et al., 2012; Chilingarian et al., 2024b). Crucially, these measurements show good agreement with predicted energy spectra (Dwyer, 2003) and emphasize the role of seed electrons from extensive air showers (EAS, Auger et al., 1939).

Free electrons are abundant in the troposphere. The height of electron maximum density, the so-called Regener–Pfotzer maximum, depends on the geomagnetic cutoff rigidity (Rc), the type of detected particles, and the strength and phase of the solar cycle. Recent measurements, confirmed by PARMA4 calculations (Sato, 2016), indicate that the maximum of the charged particle flux occurs in equatorial regions (Rc = 10-15 GV) at 16-17 km (Fig. 4 of Amrosova et al., 2023). Electric fields, induced by strong thunderstorms, transfer energy to free electrons, accelerate them, and, under special conditions, form electron-photon avalanches. In 1992, Gurevich, Milikh, and Roussel-Dupré formulated conditions for copious multiplication of electrons for each energetic seed electron injected into the strong-field region (Gurevich et al., 1992). This mechanism is now known as the Relativistic Runaway Electron Avalanche (RREA, Babich et al., 2001; Alexeenko et al., 2002). A numerical method to solve the relativistic Boltzmann equation for runaway electron beams (Symbalisty et al., 1998) helps estimate the threshold (critical) field (Babich et al., 2001; Dwyer et al., 2003) required to initiate RREA. At standard temperature and pressure in dry air at sea level, $E_{th} \approx 2.80 * n$ kV/cm, where air density (n) is relative to sea level (see the recent modest update of critical energy in Dwyer and Rassoul, 2024). This threshold field is slightly higher than the breakeven field, corresponding to the electron energy at which minimum ionization occurs. If the electrons traveled exactly along the electric field lines, this would be the threshold for runaway electron propagation and avalanche multiplication. However, the electrons' trajectories deviate due to Coulomb scattering with atomic nuclei and Møller scattering with atomic electrons, causing deviations from the ideal case. Additionally, secondary electrons from Møller scattering aren't created along the field line, so electric fields 20-30% larger are necessary for electrons to run away and initiate an avalanche.

## 2. Observation of the enhanced particle fluxes above and below thunderstorms

The Atmosphere-Space Interactions Monitor (ASIM, Neubert et al., 2019) on board the International Space Station (ISS) comprises two types of instruments: X- and gamma-ray spectrometers, and multispectral imaging arrays. The high-energy detector (HED) consists of 12 bismuth-germanium-oxide (BGO) bars coupled to a photomultiplier tube (PMT). The temporal resolution of the HED is determined by a dead time of ≈ 550 ns. The optical wide-field facilities of ASIM comprise two imaging cameras, operated at up to 12 frames per second (337 nm and 774 nm), and three high-speed photometers at 337 nm (bandwidth 5 nm), 180–230 nm, and 777.4 nm (bandwidth 4 nm), with a 100 kHz sampling rate. The absolute time tagging of the ASIM instruments is not worse than 20 ms, reaching a 10 µs accuracy when using timestamps from the lightning location networks. Thus, ASIM has unprecedented possibilities to register sequences of particle fluxes and lightning flashes in the upper atmosphere.

The Airborne Lightning Observatory (ALOFT, Ostgaard et al., 2024) in 2023 performed cutting-edge experiments measuring multiple gamma glows from tropical thunderstorms. ALOFT payload consists of five spectrometers, 30 photometers, three electric-field sensors, two radars, and two passive radiometers. Particle fluxes were monitored during flight, and gamma-ray-glowing clouds were identified in real-time to facilitate return and continuation of measurements. ALOFT detected more than five hundred gamma-ray glows during nine of the ten flights, showing that thunderclouds can emit gamma rays for hours and over huge regions (see Fig. 1 of Marisaldi et al., 2024). Glows were detected repeatedly following consecutive passages (aircraft revisited flaring thunderclouds after notification of mission physicists) over the same thundercloud system, covering $\approx 10^4$ km².

Correlated measurements of ALOFT and ASIM reveal several gamma-ray fluxes observed by ALOFT but not by ASIM during the ISS overpass. The authors (Bjørge-Engeland et al., 2024) conclude that the overwhelming gamma glow population directed to space from tropical thunderclouds is too weak to be observed from space. The source photon brightness of gamma glows is several orders of magnitude lower than what is usually attributed to TGF observed by orbiting gamma observatories. Thus, TGFs comprise only a small percentage of gamma glows copiously observed above tropical storms, as numerous TGEs are observed on the Earth's surface.

The ALOFT experiment provided groundbreaking insights, demonstrating that gamma glows and TGFs were immediately followed by Narrow Bipolar Events (NBEs), initiating significant lightning activity. These findings challenge prior hypotheses that lightning leaders supply the seed electrons for TGFs. Recent measurements from ALOFT and long-term nanosecond-scale monitoring of atmospheric discharges and particle fluxes at Aragats strongly support an alternative hypothesis: relativistic runaway electron avalanches precede and initiate the lightning leader (Chilingarian et al., 2015, 2017). The combination of gamma-ray imaging and lightning interferometry proposed by the ALOFT continuation can directly test and validate this transformative hypothesis.

The ALOFT findings, which demonstrate the role of thunderclouds as huge natural particle accelerators, confirm ground-based observations obtained by continuously monitoring particle fluxes and electric fields within large areas around Aragats Mountain (Chilingarian et al., 2022; Chilingarian et al., 2024c). Thus, the joint study of TGEs and gamma glows will enhance fundamental knowledge of high-energy physics in the atmosphere (HEPA), clarifying key issues related to particle acceleration and lightning initiation while bridging the gap between processes in the lower and upper atmosphere (Dwyer et al., 2012).

Balloon flights equipped with radiation detectors were also conducted in 2023 and are scheduled for 2025 (Pallu et al., 2023). The high-energy lightning emission network (HELEN) payloads are typically launched on a 1,200g latex weather balloon filled with helium (Helmerich, 2024). Updrafts frequently carried balloons into the storm's center. The radiation detector features a 12×12×20 mm LYSO: Ce scintillator crystal coupled with a Hamamatsu R6095 photomultiplier tube. During flights, HELEN registered gamma-ray glows lasting up to 2 minutes, with increased count rates that exceeded 10,000 counts per second.

Thus, there is a hierarchy of gamma-ray fluxes above the upper dipole in the thunderous atmosphere. Suppose an aircraft or balloon flies just above a thundercloud. In that case, millions of electrons and gamma rays will be detected, as during TGEs in the lower dipole, coming from an accelerating field 50 m above the Earth's surface. At ALOFT altitudes several kilometers above the source, RREA electrons would likely not survive, and gamma rays arrive in patches ranging from seconds to minutes. During the largest tropical thunderstorms, only a dozen gamma rays occasionally pass ≈500 km to encounter orbiting gamma-ray detectors. During the ALOFT 2024 campaign, among numerous gamma glows, no TGFs were observed by orbiting gamma observatories. Although at a smaller scale, the same hierarchy exists for TGEs: the percentage of electrons in RREA reaching the surface detectors is inversely proportional to the height of the electric field above the detectors. The greater the height, the fewer particles reach the surface, and the smaller the intensity of TGE is. If a storm sustains a strong electric field, the TGE flux persists for minutes, with particles arriving uniformly according to the Poisson law (Chilingarian et al., 2024d). Recently, it was discovered (Chilingarian et al., 2024a) that electron accelerators in thunderclouds can be astonishingly stable on a minute time scale.

## 3. Lightning flashes and particle fluxes

Historically, orbiting gamma-ray observatories (at altitudes of 400–700 km) have detected electron accelerators in thunderclouds through brief microsecond bursts of gamma rays (TGFs), which occasionally reach these observatories. However, due to their brevity and the large observational distance, TGFs provided limited insight into the physical processes of electron acceleration in the upper dipole. Recent observations of ALOFT (Marisaldi et al., 2024; Ostgaard et al., 2024), conducted 1-2 kilometers above gamma-ray sources, yield groundbreaking findings that demonstrate the existence of extensive and prolonged gamma-ray emission regions, contradicting earlier theoretical models used for 30 years to explain TGFs. These newly characterized phenomena, such as Flickering Gamma-ray Flashes (FGFs) and Glow Bursts (GBs), challenge models featuring intense distributed gamma-ray sources and open a fresh frontier in the RREA/gamma glow model similar to the RREA/TGE model. In 2012, the relation between lightning and TGF was thought to be firmly established (Dwyer et al., 2012): "TGFs are produced in what are structurally *normal IC flashes* during the period when the *initial negative polarity leader travels upward* from the main negative charge layer to the upper positive layer. This occurs during the first 5–10 ms of the lightning flash, but *distinctly after the flash initiation*. TGFs occur during the ascent of this upward leader before it reaches and expands into the upper positive charge layer. Inferred altitudes of the leader tip at the time of TGF generation have ranged from approximately 11 to 15 km.

In 2024, this relationship seemed to be smeared (Dwyer, Rassoul, 2024): "Currently, TGFs are thought to be produced inside thunderclouds during the initial stage of upward positive intra-cloud (IC) lightning. However, the *relation between lightning and TGF production is unclear,* nor is it well understood why some lightning produces TGFs while others do not. In addition to the spacecraft observations, X-rays and gamma rays are emitted by thunderclouds using both in situ and ground-based observations. These emissions often form gamma-ray "glows" lasting *seconds to minutes (thunderstorm ground enhancements* when observed from the ground). Many of these observations show energy spectra similar to TGFs that extend into the multi-MeV range, indicating similar source mechanisms. "

In 2025, it was firmly declared that "thunderstorms may generate intense gamma-ray emissions without lightning" (Dwyer, 2025).

There is plenty of evidence, starting from the first balloon and NASA's F-106 jet flights (Parks et al., 1981; McCarthy et al., 1985) to recent aircraft-based observations (Kelley et al., 2015; Kochkin et al., 2017; Ostgaard et al., 2019), that lightning suddenly stops particle fluxes in the upper atmosphere. Lightning flashes in the lower and upper atmosphere abruptly terminate particle fluxes by lowering the potential difference between charged layers in the cloud. TGEs provide extensive information on particle-lightning relations in the lower atmosphere. Multi-year joint monitoring of atmospheric discharges and particle fluxes on Aragats proves that TGEs are separated from atmospheric discharges (Chilingarian et al., 2015, 2017a). During intense particle fluxes, no nearby discharges are registered within 10 km (Chilingarian and Hovsepyan, 2022). Free electrons from extensive air showers (Auger et al., 1939) are plenty in the atmosphere at each altitude. This source of seed electrons, combined with strong electric fields, sustains electron-gamma ray avalanches below thunderclouds without any atmospheric discharges.

However, another scenario of the lightning-particle flux relation is adopted for the TGF. To justify a TGF model, it is suggested that a small-scale E-field at the lightning leader tip can be strong enough to start RREA and provide abundant seed electrons to create an extremely bright source with up to $10^{20}$ fluence (cold runaway model, Celestin & Pasko, 2011). Recent measurements and reanalysis of TGF catalogs do not support the “lightning” scenario of TGF origination (see review Chilingarian, 2024, and references therein). TGFs, measured alongside lightning flash detections by ASIM instruments, occur about 1.4 milliseconds before optical pulses begin (Skeie et al., 2022). In the multi-pulse patterns of TGFs registered by ASIM (Fig. 6.6 of Fuglestad, 2023), we can see that the first TGF detected at 18:02:25 on July 5, 2021, smoothly finished, and no optical images of the atmospheric discharges were observed. The second TGF, which occurred 2 ms later, was terminated by a lightning flash, as seen in both the abrupt termination of particle flux and the optical signal. Zhang et al. (2021) demonstrated that gamma rays are produced several milliseconds before a narrow bipolar event, which often marks lightning initiation. Analysis of four TGF catalogs from different instruments revealed that a significant proportion of TGFs lead to increased lightning activity detected in radio waves (spherics) between 150-750 ms after TGFs occur (Lindanger et al., 2022). Furthermore, in a recent paper (Gourbin and Celestin, 2024), the maximum achievable number of electrons from “cold runaway" was limited to $10^{17}$.

RREA going out from the upper and lower dipoles contains copious electrons, positrons, gamma rays, and neutrons, see Fig. 2. The fluxes directed to open space are possibly more intense than those directed to the Earth’s surface due to perhaps larger percentage over the critical value in the upper dipole than in the lower dipole. The RREA flux is less attenuated at high altitudes due to the thinning air, in contrast to the lower dipole, where the increasing air density leads to the fast attenuation of the avalanche. The overall brightness of the tropical thundercloud can fulfill the TGF electron fluence requirement to be “at least ten billion times higher than the cosmic ray particles that were passing through the thundercloud” (Dwyer, Rassoul, 2024). Without invoking an exotic source of seed electrons, the fluence of RREA in the upper atmosphere during tropical thunderstorms can reach enormous values. The PARMA4 model (Sato, 2016) estimates electron density at 15 km to be approximately $10^4$/sec $m^2$. Assuming a cloud area of $10^8$ $m^2$ and a gamma glow lasting roughly 3 hours ($\approx 10^4$ seconds), with an RREA multiplication factor of about $10^4$ (Dwyer et al., 2012; Chilingarian et al., 2022), we calculate the possible fluence of RREA electrons to reach $10^{20}$. Thus, Gamma glows lasting hours over large tropical systems can naturally amplify cosmic-ray-generated seed electrons to enormous fluences via RREA. Although ALOFT did not directly measure atmospheric electric fields, the detection of prolonged high-energy gamma-ray emissions (including FGFs and GBs) at 10–12 km altitudes implies the presence of extended electric fields exceeding the RREA threshold. Modeling these emissions reveals that fields in the 1.2–1.5 kV/m range, sustained over kilometer-scale paths, can produce the observed particle fluxes and photon energies.

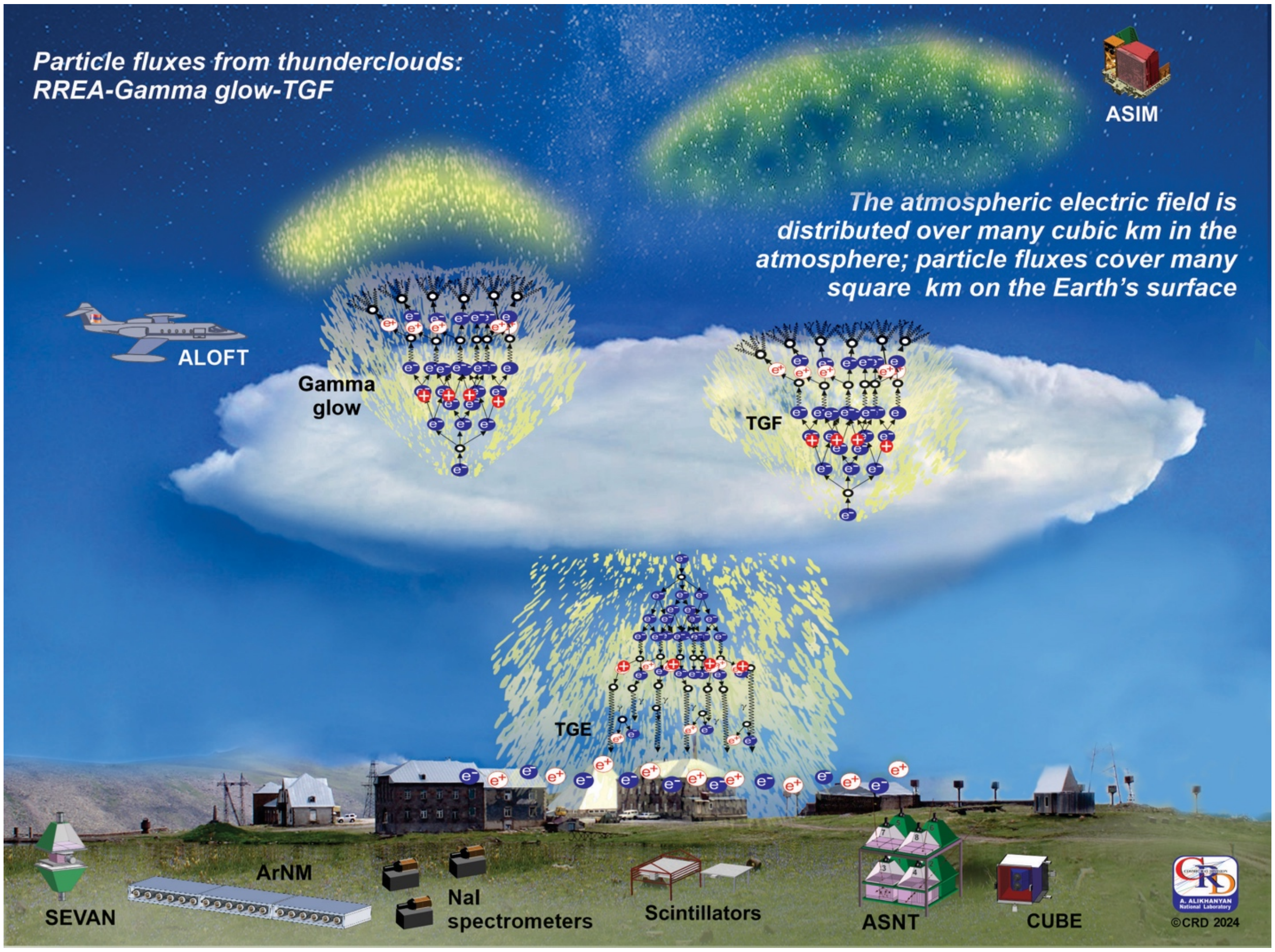

**Figure 2. RREA originates extended particle fluxes in the upper and lower atmosphere.**

## 4. Model of the DTGF origination

Wada et al. (2025) propose a model associated with converging positive and negative leaders based on detecting 31-μs gamma-ray bursts accompanied by lightning leader detection. However, their assumption that a strong electric field in the narrow gap between approaching leaders initiates RREA overlooks fundamental constraints imposed by avalanche physics and raises essential physical questions about the spatial extent of the electric field responsible for RREA.The observed DTGF duration, $\Delta t \approx 31$ μs, along with the stepped leader approach speed ($v \approx 1.8 \times 10^6$ m/s, Wada, 2025), suggests that electron acceleration continues when the gap distance is less than: $\Delta h = v \times \Delta t \approx (1.8 \times 10^6$ m/s$) \times (31 \times 10^{-6}$ s$) = \approx 56$ m.
The characteristic avalanche (e-folding) length for RREA at near-ground pressure is $\lambda \approx 100$–200 m (Celestin & Pasko, 2011; Dwyer et al., 2012). Efficient avalanche multiplication, necessary for observable DTGF production, requires several e-folding lengths to be available before the rapidly closing gap between the two leaders. This implies that if the acceleration region were confined to the leader–leader gap, RREA and significant gamma-ray emission would not be possible.

The DTGF terminates when the potential difference is neutralized by leader reconnection. In this model, the upward positive leader does not initiate the DTGF; instead, it terminates it by neutralizing the potential difference between the negative leader and Earth, explaining the abrupt cessation of the gamma-ray burst. The acceleration process is thus sustained in the full vertical leader–ground gap, rather than in the narrow region between the approaching leader tips, see Figure 3.

Multiyear DTGF observations in Uchinada, Japan, are also linked to negative cloud-to-ground lightning, demonstrating that gamma-ray fluxes span a vast area of at least 100 $km^2$ around the lightning flash (see Fig. 1 of Ortberg et al., 2024). This raises a fundamental question: how can a small negative leader tip, only a few $cm^2$ in size, influence atmospheric electric fields over 5 kilometers to initiate RREA across extensive cloud volumes?

In contrast, the downward propagation of a negative leader in a pre-critical dipole field between the main negative (MN) layer and image charges near the ground can induce a non-local reorganization of the AEF. Thus, the negative leader acts not just as a direct accelerator but as a trigger for large-scale field transformation, displacing equipotential surfaces and enhancing preexisting subcritical fields over a vast volume. As the negative leader descends, it effectively 'pulls down' the potential of the MN layer, compressing the equipotential surfaces below it. This deformation increases the local electric field between the MN layer and the ground, transforming previously subcritical regions into zones that extend 800 m vertically and 10 km horizontally, capable of initiating RREA. This mechanism, demonstrated by the detection of hundreds of TGEs worldwide (Chilingarian et al., 2025), allows extensive areas, spanning many square kilometers, to experience a deepening of negative AEF from modest levels to supercritical values. Thus, the DTGFs observed in Uchinada receive a natural explanation within the large-scale extension of AEF in the lower dipole, rather than assuming an enigmatic source of $10^{19}$ electrons above 1.5 km from the surface.

This model aligns with DTGF observations at Uchinada, recent gamma glows (Marisaldi et al. 2024; Ostgaard et al., 2024), and multi-year observations of thunderstorm ground enhancements (TGEs) by the SEVAN network. (Chilingarian et al., 2010; 2011). These observations reveal that high-energy particle bursts can affect regions covering many square kilometers before lightning occurs.

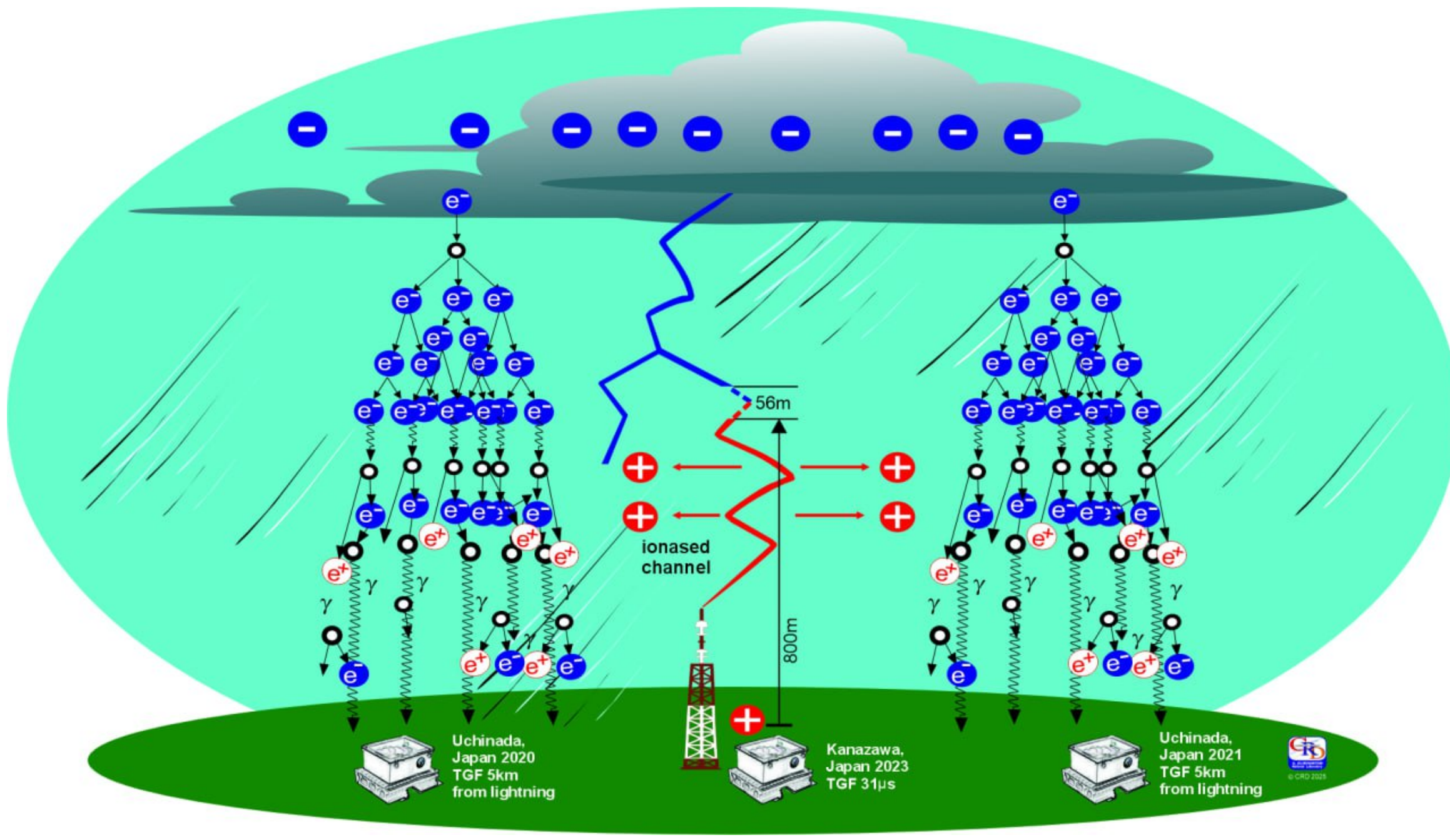

**Figure 3. Revised DTGF model. Approaching lightning leaders (center) induce non-local reorganization of the atmospheric electric field (AEF), transforming previously subcritical regions into RREA-capable zones. RREA regions are shown at distances of ~5 km from the lightning channel, consistent with observations in Uchinada (2020–2021).**
**Blue – downward negative leader; red – upward positive leader.**

## 5. Conclusion

Charge separation in thunderclouds, driven by updrafts of warm air and interactions among hydrometeors, generates oppositely directed dipoles within the cloud. The oppositely directed atmospheric electric field (AEF) in upper and lower dipoles accelerates free electrons toward open space and the Earth's surface. Free electrons are abundant in the troposphere due to extensive air showers (EASs, Auger, 1950). Electric fields induced by strong thunderstorms transfer energy to these electrons, accelerating them and, under specific conditions, forming electron-photon avalanches. These avalanches propagate through large atmospheric volumes, extending across extensive areas upon reaching the Earth's surface, and significantly increasing natural gamma radiation (NGR), which has critical implications for radiation safety and aviation. These enhancements can last from seconds to tens of minutes. The ionized channels created by these avalanches provide pathways for lightning leaders to propagate toward the ground.

Relativistic Runaway Electron Avalanches (RREA) are central to understanding a spectrum of high-energy atmospheric phenomena, including Thunderstorm Ground Enhancements (TGEs), gamma-ray glows, upward and downward terrestrial gamma flashes.

Recognizing the same origin for microsecond TGFs, minute-long gamma glows, and TGEs marks a step towards accepting RREA and EAS as a universal physical process, responsible for the enhanced particle fluxes in the lower and upper atmosphere.

The following terminology pertains to the atmospheric fluxes of elementary particles produced during thunderstorms:
Terrestrial gamma-ray flashes (TGFs) are brief bursts of gamma radiation detected by orbiting gamma-ray observatories. They last for tens of microseconds, originate from thunderstorms at altitudes between 10 and 15 kilometers in equatorial regions, and are identified by gamma-ray observatories positioned 400 to 700 kilometers above the source.

- TGEs (thunderstorm ground enhancements) refer to intense and prolonged particle fluxes observed on the Earth's surface, lasting from seconds up to several tens of minutes. These fluxes originate from the same RREA process; the accelerating electric fields are near the detectors (sometimes 25-100m), allowing for a detailed study of the energy spectra of electrons and gamma rays, as well as the charge structures of thunderclouds.
- Gamma glows are bursts of gamma radiation detected in the atmosphere by instruments carried on balloons or aircraft. These events can last tens of seconds to several minutes and typically conclude with a lightning strike.
- The gamma-ray enhancements observed at the Earth's surface are sometimes also called gamma glows because thunderclouds, due to their high altitude, exclusively detect gamma rays and no other types of radiation particles (Wada et al., 2021).

- The negative leader initiates downward TGFs, which act not just as a direct accelerator but as a trigger for large-scale field transformation, displacing equipotential surfaces and enhancing preexisting subcritical fields over vast thunderstorm volumes.

The primary mechanism driving TGFs, DTGFs, gamma glows, and TGEs is the RREA, which utilizes the ambient population of EAS-free electrons as seeds. This process originates intense electron-gamma-ray avalanches that can overpass the background radiation hundreds of times and accelerate seed electrons to 50-60 MeV energies. Simultaneous measurements of TGE electrons and gamma-rays confirmed RREA as a viable electron accelerator covering many $km^2$ areas on the Earth's surface.
However, a disconnect persists in the TGF research community, where the physical processes occurring in the upper atmosphere are often considered entirely different from those in the lower atmosphere. For instance, a recent TGF paper (Østgaard et al., 2024) states, "Two different hard-radiation phenomena are known to originate from thunderclouds: terrestrial gamma-ray flashes (TGFs) and gamma-ray glows." It is perplexing that such a distinction is made, given that the same physical mechanism—RREA—operates equally in the lower and upper atmospheres. This oversight neglects the vast wealth of TGE data. Nearly 1,000 TGEs have been observed on mountaintops across Eastern Europe, Japan, Russia, Germany, and Armenia (Chilingarian et al., 2025). Combined with simulations, these observations have provided a detailed understanding of RREA development and cloud charge structures in the lower atmosphere.

Gamma glows, originating in upper atmospheric dipoles, exhibit different characteristics from TGEs. The high altitude, thinner air, create distinct conditions for RREA development. Therefore, reserving the term "gamma glow" for particle fluxes observed in the upper atmosphere and "TGE" for particles observed on the Earth's surface is appropriate. Despite their different locations, these two phenomena—both powered by RREA—share similarities and would likely resemble each other if measured at the same distance from the source. Recent findings from the ALOFT mission support this conclusion (Marisaldi et al., 2024).
TGF was discovered in the 1990s to describe brief bursts of gamma rays detected by satellites. TGF models propose that seed electrons originate not from cosmic rays but from streamers or lightning leaders. TGFs are rather an exotic phenomenon. Detecting microsecond-duration bursts of gamma rays by satellites positioned 400-700 kilometers from their source provides limited information for physical analysis compared to the wealth of data gathered from TGEs and gamma glows. Moreover, the spatial resolution of TGF measurements is inherently lower due to the great distance between the "accelerator" and the detector, as opposed to TGE and gamma glow detectors.

Broad scientific understanding requires incorporating the various experimental aspects of RREA phenomena that give rise to TGFs, TGEs, and gamma glows. Neglecting any part of this multidisciplinary research would lead to erroneous conclusions and narrow physical interpretations. The statements like “The phenomenology of TGFs and glows has been poorly understood” and questions like “Are glows necessarily precursors to TGFs? Is there a competition between TGFs, glows, and regular lightning for electric field energy?” (Lang et al., 2025) that continuously appear in the papers will remain unanswered without broad collaboration and a revision of the entire HEPA results. I strongly advocate for a community-wide revision of the HEPA results obtained in space and on the Earth’s surface, and for clarifying the terms used to describe these phenomena to provide a unified platform for discussing and united HEPA research. This will help prevent further confusion and ensure proper citation policies in publications.

**Reference**

Alexeen.ko V.V, Khaerdin.ov N.S, Lidvansky A.S, et al. (2002). Transient variations of secondary cosmic rays due to atmospheric electric field and evidence for pre-lightning particle acceleration. Phys Lett A 301:299–306.

Ambrozová I., Kákona M., Dvorák R., et al. (2023) Latitudinal effect on the position of Regener–Pfotzer maximum investigated by balloon flight HEMERA 2019 in Sweden and balloon flights FIK in Czechia, Radiation Protection Dosimetry 199(15–16), 2041. doi.org/10.1093/rpd/ncac299

Auger P., Ehrenfest P., Maze, R., et al.: 1939, Extensive Cosmic-Ray Showers, Reviews of Modern Physics, 11, 3–4, 288–291. doi.org/10.1103/RevModPhys.11.288

Babich, L. P., Donskoy, E. N., Kutsyk, I. M., & Kudryavtsev, A. Y. (2001). *Comparison of relativistic runaway electron avalanche rates obtained from Monte Carlo simulations and kinetic equation solution*. IEEE Transactions on Plasma Science, 29(3), 430-438.

Bjørge-Engeland, I.,Østgaard, N., Sarria, D., et al.(2024).Evidence of a New population of weak Terrestrial Gamma-Ray flashes observed from aircraft Altitude, Geophysical Research Letters,51,e2024GL110395. doi.org/10.1029/2024GL110395

Celestin, S., & Pasko, V. P. (2011). Energy and fluxes of thermal runaway electrons produced by the exponential growth of streamers during the stepping of lightning leaders and in transient luminous events. Journal of Geophysical Research (Space Physics), 116, 3315. doi: 10.1029/2010JA016260

Chilingarian A., A. Daryan, K. Arakelyan, et al. (2010) Ground-based observations of Thunderstorm-correlated fluxes of high-energy electrons, gamma rays, and neutrons, Phys. Rev.. D 82, 043009. doi: 10.1103/PhysRevD.82.043009

Chilingarian A, Hovsepyan G., and Hovhannisyan A. (2011) Particle bursts from thunderclouds: Natural particle accelerators above our heads, Phys. Rev.. D 83, 062001. doi.org/10.1103/PhysRevD.83.062001

Chilingarian A., Bostanjyan N., Vanyan L., (2012) Neutron bursts associated with thunderstorms, Phys. Rev.. D 85, 085017.

Chilingarian A., Hovsepyan G., Khanikyanc Y., Reymers A., and Soghomonyan S. (2015) Lightning origination and thunderstorm ground enhancements terminated by the lightning flash, Europhysics Letters, 110, 49001. Doi: 10.1209/0295-5075/110/49001

Chilingarian A., Chilingaryan S., Karapetyan T., Lev Kozliner, Yeghia Khanikyants, Gagik Hovsepyan, David Pokhsraryan & Suren Soghomonyan (2017) On the initiation of lightning in thunderclouds, Scientific Reports 7, Article number: 1371, doi:10.1038/s41598-017-01288-0.

Chilingarian A., Hovsepyan G., Karapetyan T., et al. (2022), Development of the relativistic runaway avalanches in the lower atmosphere above mountain altitudes, EPL, 139, 50001, https://doi.org/10.1209/0295-5075/ac8763

Chilingarian A., Hovsepyan G. (2022), The synergy of the cosmic ray and high energy atmospheric physics: Particle bursts observed by arrays of particle detectors, New Astronomy, 97, 101871.

Chilingarian A., (2024) Extensive air showers and atmospheric electric fields. Synergy of space and atmospheric particle accelerators, Advances in Space Research 74, 4388. doi.org/10.1016/j.asr.2024.03.013

Chilingarian A., Sargsyan B., Karapetyan T., et al. (2024a) Extreme thunderstorm ground enhancements registered on Aragats in 2023, Physical Review D 110, 063043a

Chilingarian A., Sargsyan B. (2024b) Atmospheric positron flux modulation during thunderstorms, Physical Review D 109, 062003. doi:10.1103/PhysRevD.109.062003

Chilingarian A., Karapetyan T., Sargsyan B., Khanikyanc Y., and Chilingaryan S. (2024c) Measurements of Particle Fluxes, Electric Fields, and Lightning Occurrences at Aragats Space-Environmental Center (ASEC), Pure and Applied Geophysics 181, 1963. https://doi.org/10.1007/s00024-024-03481-5

Chilingarian A., Pokhsraryan D., Zagumennov F., Zazyan M. (2024d) Space-temporal structure of the thunderstorm ground enhancements (TGEs), Physics Open 18, 100202. doi.org/10.1016/j.physo.2023.100202

Chilingarian A., Karapetyan T., Sargsyan B., Knapp J., Walter M., Rehm T.(2024e) Energy spectra of the first TGE observed on Zugspitze by the SEVAN light detector compared with the energetic TGE observed on Aragats, Astroparticle Physics 156, 02924. doi: 10.1016/j.astropartphys.2024.102924

Chilingarian, A., Williams, E., Hovsepyan, G., & Mkrtchyan, H. (2025). Why Schonland failed in his search for runaway electrons from thunderstorms. *Journal of Geophysical Research: Atmospheres*, *130*, e2024JD042350. doi.org/10.1029/ 2024JD042350

Chum, R. Langer, J. Baše, M. Kollárik, I. Strhárský, G. Diendorfer, J. Rusz (2020) Significant enhancements of secondary cosmic rays and electric field at high mountain peak during thunderstorms, Earth Planets Space 72, 28. doi.org/10.1186/s40623-020-01155-9

Dwyer J.R. (2003) A fundamental limit on electric fields in air, Geophys. Res. Lett. 30, 2055. doi.org/10.1029/2003GL017781

Dwyer J.R, Smith D.M., Cummer S.A. (2012) High-Energy Atmospheric Physics: Terrestrial Gamma-Ray Flashes and Related Phenomena, Space Sci Rev 173, 133. DOI 10.1007/s11214-012-9894-0

Dwyer G.R., Rassoul H.K. (2024) High energetic radiation from thunderstorms and lightning, in Lightning Electromagnetics, IET series Volume 1, 365.

Dwyer,J.R.(2025). Energetic particles are produced by thunderstorm electric fields. Journal of Geophysical Research: Atmospheres,130,e2024JD042193. https://doi.org/10.1029/2024JD042193

Fishman G. J., Bhat P.N., Mallozzi R., et al. (1994) Discovery of Intense Gamma-Ray Flashes of Atmospheric Origin, Science, 264, 1313. doi/science.264.5163.1313

Fuglestad A. N., Multi Pulse Terrestrial Gamma-ray Flashes and optical pulses of lightning observed by ASIM, Master Thesis in Space Physics, University of Bergen https://bora.uib.no/bora-xmlui/bitstream/handle/11250/3073302/Multi-Pulse-Terrestrial-Gamma-ray-Flashes-and-optical-pulses-of-lightning-observed-by-ASIM-print-version.pdf?sequence=3&isAllowed=y

Gourbin, P., Celestin, S. (2024). On the self-quenching of relativistic runaway, electron avalanches producing terrestrial gamma-ray flashes: Geophysical Research Letters, 51(10), e2023GL107488.

Helmerich, C., McKinney, T., Cavanaugh, E., & Dangelo, S. (2024) TGFs, gamma-Ray glows, and direct lightning strike radiation is observed during a single flight of a balloon-borne gamma-ray spectrometer, Earth and Space Science, 11, e2023EA003317. doi.org/10.1029/2023EA003317

Gurevich, G., Milikh, R., Roussel-Dupre (1992) Runaway electron mechanism of air breakdown and preconditioning during a thunderstorm, Physics Letters A 165 (5), 463.

Kelley N.A., Smith D.M., Dwyer J.R., et al. (2015) Relativistic electron avalanches as a thunderstorm discharge competing with lightning, Nature Communications, 6 (7845). doi.org/10.1038/ncomms8845

Kochkin P., Van Deursen P., Marisaldi M., et al. (2017) In-flight observation of gamma-ray glows by ILDAS. JGR, Atmos 122:12801–12811. doi: 10.1002/2017JD027405

Kuettner, J. (1950). The electrical and meteorological conditions inside thunderclouds. *J. Meteorol.*, *7*, 322–332. doi.org/10.1175/1520-0469(1950)007<0322:TEAMCI>2.0.CO;2

Lang T.J., Østgaard N., Marisaldi M. et al., Hunting for Gamma Rays above Thunderstorms: The ALOFT Campaign, Bulletin of the American Meteorological Society. doi.org/10.1175/BAMS-D-24-0060.1

Leppänen, A-P., K. Peräjärvi1, J. Paatero, J. Joutsenvaara, A. Hannula, A. Hepoaho, P. Holm, J. Kärkkäinen (2024), Thunderstorm ground enhancements in Finland: observations using spectroscopic radiation detectors, Acta Geophysica, https://doi.org/10.1007/s11600-024-01495-0

Lindanger, A., Marisaldi, M., Sarria, D., Østgaard et al. (2021) Spectral analysis of individual terrestrial gamma-ray flashes detected by ASIM. Journal of Geophysical Research: Atmospheres,126, e2021JD035347. doi.org/10.1029/2021JD035347

Marisaldi M., Østgaard N., Mezentsev A., et al. (2024). Highly dynamic gamma-ray emissions are common in tropical thunderclouds, Nature 634, 57.

Neubert, T., Østgaard, N., Reglero, V., Blanc, E., Chanrion, O., Oxborrow, C. A., et al. (2019) The ASIM mission on the International Space Station. *Space Science Reviews*, *215*(2), 26. doi.org/10.1007/s11214-019-0592-z

MacGorman, D. R. and Rust, W. D., The Electrical Nature of Storms, Oxford University Press, New York, NY, 1998.

McCarthy, M., and Parks, G. (1985). Further observations of X-rays inside thunderstorms, Geophysical Research Letters, 12, 393. doi.org/10.1029/GL012i006p00393

Mailyan B.G., Briggs M. S., Cramer E. S., et al. (2016) The spectroscopy of individual terrestrial gamma-ray flashes: Constraining the source properties, J. Geophys. Res. Space Physics, 121, 11,346–11,363. doi.org/10.1002/2016JA022702

Ortberg, J., Smith, D. M., Kamogawa et al. (2024). Two laterally distant TGFs from negative cloud-to-ground strokes in Uchinada, Japan. Journal of Geophysical Research: Atmospheres, 129, e2023JD039020. https://doi. org/10.1029/2023JD039020

Ostgaard, N., Christian, H. J.Grove, J. E., Sarria, D., Mezentsev, A., Kochkin, P., et al. (2019) Gamma-ray glow observations at 20-km altitude. Journal of Geophysical Research: Atmospheres, 124, 7236–7254. doi.org/10.1029/2019JD030312.

Østgaard N., Mezentsev A., Marisaldi M., et al. 2024, Flickering gamma-ray flashes, the missing link between gamma glows and TGFs, Nature 634, 53.

Parks, G.K., Mauk, B.H., Spiger, R., Chin, J. (1981) X-ray enhancements detected during thunderstorm and lightning activities. Geophys. Res. Lett. 8, 1176–1179,

doi.org/10.1029/GL0 08i011p01176.

Pallu, M., Celestin, S., Hazem, Y., Trompier, F., & Patton, G. (2023). XStorm: A new gamma ray spectrometer for detecting close proximity gamma ray glows and TGFs. *Journal of Geophysical Research: Atmospheres*, *128*, e2023JD039180. https://doi. org/10.1029/2023JD039180

Regener, E., and Pfotzer, G. (1934). Intensity of cosmic ultra-radiation in the stratosphere with the tube counter. Nature 134, 325.

Roussel-Dupré R., Symbalisty E., Taranenko Y., Yukhimuk V. (1998) Simulations of high-altitude discharges initiated by runaway breakdown, J. Atmos. Sol.-Terr. Phys. 60, 917–940. doi.org/10.1016/S1364-6826(98)00028-5

Sato, T. Analytical model for estimating the zenith angle dependence of terrestrial cosmic ray fluxes. *PLoS ONE* **2016**, *11*, e0160390.

Sommer M., Czakoj T., Ambrožová[1]I., et al., Exploring the Potential of Aerial and Balloon–Based Observations in the Study of Terrestrial Gamma-Ray Flashes, preprint. doi.org/10.5194/egusphere-2024-2789

Symbalisty, E. M. D., R. A. Roussel-Dupré, and V. A. Yukhimuk (1998), Finite volume solution of the relativistic Boltzmann equation for electron avalanche studies, IEEE Trans. Plasma Sci., 26, 1575–1582.

Zhang, H., Lu, G., Liu, F., Xiong, S., Ahmad, M. R., Yi, Q., et al. (2021) On the Terrestrial Gamma-ray Flashes preceding narrow bipolar events. Geophysical Research Letters, 48, e2020GL092160. Doi. org/10.1029/2020GL092160

Ursi A., Virgili D., Campana R., et al. (1924) Detection of an Intense Positron Burst During a Summer Thunderstorm on Mt. Etna, AGU fall meeting AE01-04

Wada, Y., Enoto, T., Kubo, M., Nakazawa, K., Shinoda, T., Yonetoku, D., et al. (2021). Meteorological aspects of gamma-ray glows in winter thunderstorms. Geophys. Res. Lett., 48(7). https://doi.org/10.1029/2020gl091910

Wada, Y., Morimoto T., Wu T., et al. (2025). Inverse Compton gamma rays are produced before lightning. Science Advances, 11, eads6906.